\documentclass[runningheads]{llncs}
\usepackage{graphicx,amsmath,amssymb,booktabs,multirow,siunitx,etoolbox,hyperref,soul,mathtools,authblk}

\DeclareMathOperator*{\argmin}{arg\,min}
\newcommand{\norm}[1]{\left\lVert#1\right\rVert}

\usepackage{makecell}
\usepackage[export]{adjustbox}

\begin{document}

\title{Holistic Multi-Slice Framework for Dynamic Simultaneous Multi-Slice MRI Reconstruction}

\titlerunning{Holistic Multi-Slice Framework for Dynamic SMS}


\author{Daniel H. Pak\inst{1}\footnote{This work was done during the internship at United Imaging Intelligence} \and Xiao Chen\inst{2} \and Eric Z. Chen\inst{2} \and Yikang Liu\inst{2} \and Terrence Chen\inst{2} \and Shanhui Sun\inst{2}}

\authorrunning{D. H. Pak et al.}

\institute{Biomedical Engineering, Yale University, New Haven, CT, USA \and United Imaging Intelligence, Cambridge, MA, USA}

\maketitle              

\begin{abstract}

Dynamic Magnetic Resonance Imaging (dMRI) is widely used to assess various cardiac conditions such as cardiac motion and blood flow. To accelerate MR acquisition, techniques such as undersampling and Simultaneous Multi-Slice (SMS) are often used. Special reconstruction algorithms are needed to reconstruct multiple SMS image slices from the entangled information. Deep learning (DL)-based methods have shown promising results for single-slice MR reconstruction, but the addition of SMS acceleration raises unique challenges due to the composite k-space signals and the resulting images with strong inter-slice artifacts. Furthermore, many dMRI applications lack sufficient data for training reconstruction neural networks. In this study, we propose a novel DL-based framework for dynamic SMS reconstruction. Our main contributions are 1) a combination of data transformation steps and network design that effectively leverages the unique characteristics of undersampled dynamic SMS data, and 2) an MR physics-guided transfer learning strategy that addresses the data scarcity issue. Thorough comparisons with multiple baseline methods illustrate the strengths of our proposed methods.



\end{abstract}

\section{Introduction}
Magnetic Resonance Imaging (MRI), especially dynamic MRI (dMRI), is widely used for cardiac applications because it is safe and provides time-series images with great soft tissue contrast \cite{earls2002cardiac}. Cine cardiac MRI primarily captures the cardiac motion, whereas first-pass perfusion (FPP) captures the contrast passage through the heart to estimate blood flow. Both techniques 
are crucial in assessing various conditions such as coronary artery disease, arrhythmia, and valve regurgitation \cite{garg2020assessment,white1987left,wilke1999magnetic}. 

\begin{figure}[t] 
  \centering
  \centerline{\includegraphics[width=11.5cm]{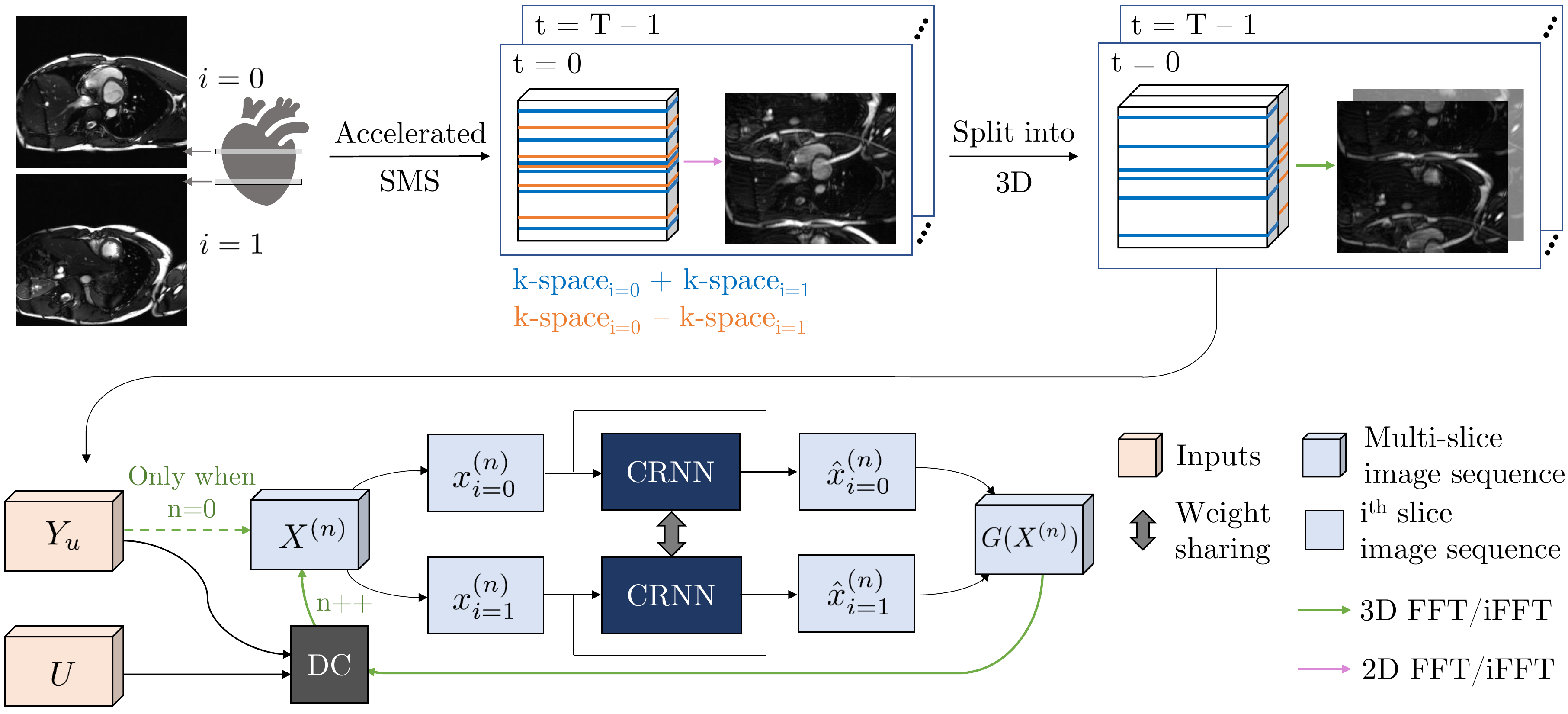}}
  \caption{The overall pipeline for 
  for a 2-slice SMS ($i \in \{0, 1\}$). The method can be applied to an arbitrary number of SMS slices.
  }
  \label{fig:recon_outline}
\end{figure}

Despite its strengths, MRI acquisition is inherently slow, which makes dMRI susceptible to motion artifacts, limited spatial coverage, and prolonged scans \cite{axel2016accelerated}. Many acceleration strategies have been proposed to manipulate the acquisition in k-space, a.k.a. the spatial frequency domain from which MR images are reconstructed.
One popular single-slice acceleration strategy is undersampling, where less k-space data is acquired in the imaging plane
\cite{zbontar2018fastmri}. Another popular acceleration technique is Simultaneous Multi-Slice (SMS) \cite{barth2016simultaneous}, where multiple slices are excited simultaneously to generate composite signals (Fig. \ref{fig:recon_outline}).
One benefit of SMS is that inter-slice acceleration comes with little to no cost of signal-to-noise ratio, unlike intra-slice acceleration by undersampling. Studies have shown that great acceleration and spatial coverage can be achieved by combining SMS with undersampling \cite{barth2016simultaneous}.

The success of these acceleration techniques relies heavily on reconstruction algorithms to recover clinically acceptable image quality from undersampled data. Direct inverse Fourier transform on zero-filled undersampled k-spaces produces images with aliasing artifacts due to the violation of the Nyquist Theorem. For undersampled SMS, the resulting images have additional inter-slice artifacts, characterized by overlapping contents from all excited slices (Fig. \ref{fig:recon_outline}). Parallel imaging (PI) is commonly used to reduce these artifacts when the acceleration rate is low \cite{deshmane2012parallel}, but it is suboptimal for cardiac MRI (CMR) reconstruction due to insufficient coil sensitivity along the slice direction \cite{barth2016simultaneous}. Compressed sensing (CS) can handle both inter- and intra-slice acceleration, but relies on manually crafted features and has low reconstruction speed due to the iterative optimization process \cite{lustig2008compressed}.



Recently, deep learning-based methods have demonstrated superior reconstruction performance over PI and CS \cite{lee2018deep,lundervold2019overview}. Previous studies on single-slice static image reconstruction have utilized convolutional neural networks (CNNs) for single forward-pass reconstruction \cite{hyun2018deep,zbontar2018fastmri} or iterative denoising \cite{schlemper2017deep,hammernik2018learning}. Others have proposed convolutional recurrent neural networks (CRNNs) for dynamic MR reconstruction \cite{chen2020real,qin2018convolutional} and self-supervised methods for static SMS reconstruction \cite{demirel2021improved}. However, none of these works directly address dynamic SMS reconstruction. The only existing deep learning-based dynamic SMS reconstruction methods mostly rely on PI to first separate the SMS slices and then denoise each slice separately using CNNs \cite{le2021deep,wang2021deep}. In contrast, our proposed framework handles slice separation and aliasing directly within the model, thus enabling holistic deep learning from the original undersampled SMS k-space data to the final multi-slice reconstructed images.



Most deep learning-based reconstruction methods require a paired dataset of undersampled and fully sampled k-spaces. However, raw k-space data are difficult to obtain because MR acquisition protocols generally discard the raw data after generating the post-processed DICOM images \cite{zbontar2018fastmri}. In addition, some CMR applications such as FPP require a contrast agent injection, which limits the subject pool and further exacerbates the difficulty in data acquisition. This results in a major data shortage problem for deep learning-based FPP reconstruction. Thus, we propose a transfer learning strategy from cine CMR, which is performed routinely in the clinic without contrast injection and has more data available. 


In this study, we present deep learning-based reconstruction methods for dynamic SMS MR reconstruction. Our main contributions are 1) a novel approach that addresses the inter-slice interactions of dynamic SMS k-spaces and images, and 2) an MR physics-based transfer learning strategy that alleviates the FPP data scarcity issue. Thorough comparisons with multiple baseline methods demonstrate the strengths of our proposed methods.

\section{Methods}





\subsection{Problem Formulation}

Let $x_i \in \mathbb{C}^{a \times b \times t}$ be the image sequence of the $i^{th}$ SMS slice, where $a$, $b$, and $t$, denote the 2D spatial and time dimensions. A simple SMS acquisition has the k-space data as $y = \sum_{i=0}^{M-1}y_i = \sum_{i=0}^{M-1}F_{2D}(x_i)$, where $F_{2D}$ is the 2D Fourier transform operator along $a$ and $b$, and $M$ is the number of SMS slices. Note that $y$ here does not have the slice dimension and the aim of the reconstruction is to recover $x_i$ from $y$. Using phase modulation during data acquisition \cite{barth2016simultaneous}, each k-space readout line from slice $i$ has an accumulating phase of $2\pi i/M$. The SMS acquisition can then be viewed as acquiring data in a 3D k-space  (proof in supplement):
\begin{subequations}
\begin{align}
    y_k &= \sum_{i=0}^{M-1}F_{2D}(x_{i} \cdot e^{-j\frac{2\pi}{M}ik})  \\
    Y &= F_{3D}(X) \label{eq:F3D}
\end{align}
\end{subequations}

\noindent $y_k$ is an apparent and additional $k^{th}$ band in k-space. Eq. \ref{eq:F3D} is a result of the Fourier transform definition, where $F_{3D}$ is the 3D Fourier transform operator along $a$, $b$, and $i$, and $X$ and $Y$ are concatenations along slice dimensions of $x_i$ and $y_k$, respectively (Fig.\ref{fig:recon_outline}). For brevity, $F \coloneqq F_{3D}$ henceforth. The SMS reconstruction problem can then be treated as a 3D reconstruction problem:
\begin{equation} \label{eq:optimization}
\argmin_{X} \mathcal{R}(X) + \lambda \norm{Y_u - U \odot F(X)}_2^2
\end{equation}
where $\mathcal{R}$ is the regularization on $X$, $U$ is the binary undersampling mask, and $Y_u$ is the undersampled SMS k-space data in the 3D format. Following the derivation in \cite{qin2018convolutional,schlemper2017deep}, the DL-based solution to Eq.\ref{eq:optimization} can be formulated as iterating a neural network forward pass $G(\cdot)$ to denoise the current image estimation and a data consistency (DC) step (Fig.\ref{fig:recon_outline}): 
\begin{equation}
     X^{(n+1)} = F^{-1}(Y_u + (1-U) \odot F(G(X^{(n)})))
\end{equation}

\subsection{DL-based Reconstruction Pipeline for dynamic SMS}


For dynamic SMS CMR reconstruction, previous works reconstructed each slice separately and used a 3D U-net on the 2D + time images stacked in the time direction \cite{le2021deep,wang2021deep}. The inter-slice artifacts are handled during preprocessing with PI techniques, which relies heavily on the quality of coil sensitivity maps (CSM) \cite{deshmane2012parallel}. For CMR applications, the CSMs do not have good configuration along the slice direction, which may lead to deteriorated image quality after PI-based reconstruction \cite{barth2016simultaneous}. This is why most current dynamic CMR acquisitions are performed one 2D slice at a time, where the process is repeated for multiple time frames and slices to capture the dynamics of the whole heart \cite{axel2016accelerated}.

In SMS, slices are usually placed far apart to avoid slice cross-talk (Fig. \ref{fig:recon_outline}). Even so, the undersampled SMS images still have strong inter-slice correlations because each slice contains aliasing artifacts from all other slices (Fig. \ref{fig:recon_outline}). To leverage this characteristic, we propose a holistic framework that processes all SMS slices together in the reconstruction pipeline. As shown in Fig. \ref{fig:recon_outline}, all slices' k-space data $Y_u$ is directly used in the network without incorporating any coil sensitivity map calculations. The initial undersampled images are obtained via an inverse 3D Fourier transform, and the time series images of each slice are processed using a weight-sharing CRNN to leverage the correlation between the SMS slices. We hypothesized that weight-sharing would be beneficial because it would allow CRNNs to perform local 2D + time denoising, and the repeated content from the inter-slice artifacts would help identify similar local 2D features between all slices to either remove or strengthen in each slice. This is different from say, a 3D convolution, which would combine inter-slice information without considering the inter-slice distance and the translations in inter-slice artifacts.









\begin{figure}[t]
  \centering
  \centerline{\includegraphics[width=11.5cm]{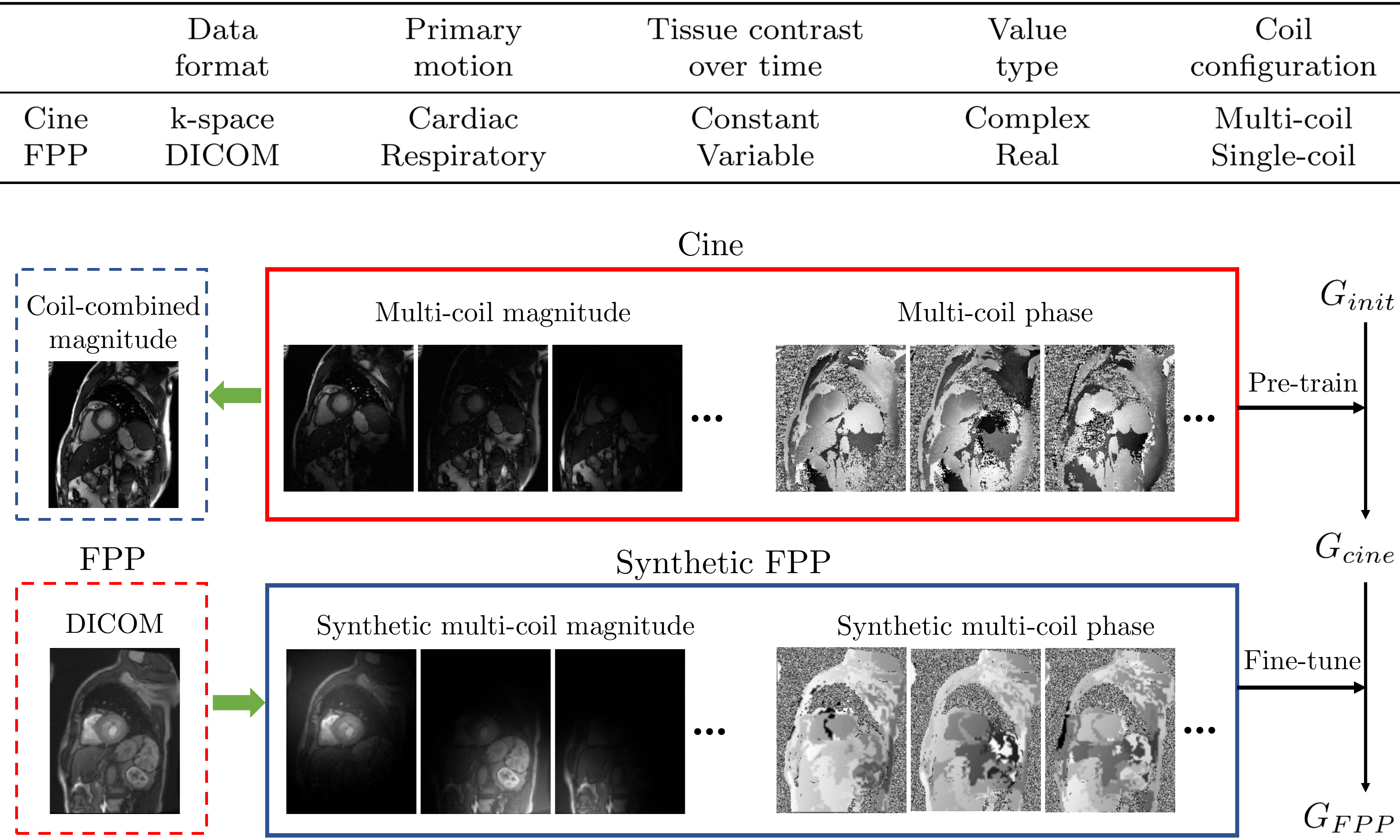}}  
  \caption{Our proposed transfer learning strategy for FPP reconstruction. Red: original data. Blue: data derived from the original data. Only the solid boxes are used for training. One representative image slice is shown for each dataset.}
  \label{fig:transfer_method}
\end{figure}


\subsection{Transfer Learning Strategy for FPP Reconstruction}

Our transfer learning strategy is briefly illustrated in Fig. \ref{fig:transfer_method}. First, we establish two important assumptions: (1) we have a large cine k-space dataset, and (2) we have a much smaller FPP DICOM dataset. Given these conditions, our strategy is to first fully train a model using the cine data, and subsequently fine-tune it using a synthetic FPP k-space dataset, which we generate using the FPP DICOM data and MR principles.

Of the four key differences between the original cine and FPP data (Fig. \ref{fig:transfer_method}), the image-based information such as primary motion and changes in tissue contrast are attributes contained within the DICOM data. Thus, the crux of the problem is minimizing the domain gap caused by the different value types and coil configurations (Fig. \ref{fig:transfer_method}, more evidence provided in the supplement).



Based on the complex number definition, $C = R \cdot \exp(j\phi)$, we need suitable phase information, $\phi$, to convert DICOM into synthetic complex-valued data. Also, based on the multi-coil definition $X_c = X \odot CSM_c$ for the $c^{th}$ coil image, we need reasonable estimation of the coil sensitivity maps (CSM) to generate synthetic multi-coil data (Fig. \ref{fig:transfer_method}).

For synthetic phase, previous MR reconstruction studies have utilized simple sinusoidal gratings \cite{dar2020transfer,knoll2019assessment}. We improved our estimation of phase for dynamic SMS CMR using a few key observations: phase varies significantly across different coils and structural boundaries, but it does not vary significantly across time and different SMS slices of the same coil. We generated phases that satisfy all of these conditions via the following procedure:

\begin{enumerate}
    \item For each SMS slice of time-series DICOM,  apply 3D k-means clustering across spatial and time dimensions.
    \item For one SMS slice, assign to each cluster a value randomly sampled from $\mathcal{N}(0, \pi/3)$ (99.7\% of values within $-\pi$ and $\pi$). Repeat for the number of desired coils.
    \item For the remaining SMS slices, assign to each cluster (for the corresponding coil) a value based on majority voting using values from step 2.
    \item Add to a sinusoidal grating, which is random across different coils and patients and constant across time and SMS slices.
\end{enumerate}

Then, we generated synthetic CSM's using two different approaches. (1) We assumed a circular arrangement of coils and greater sensitivity near each coil (sensitivity $:= 1/d^3$, where $d$ is the distance from each coil). (2) We used random 2D Gaussian blobs to add variety in the shape and location of the CSMs. Here, the random variables are the center position, primary eigenvalue, eigenvalue ratio, and rotation. Finally, we applied the described synthetic phases and CSMs to generate multi-coil k-space data from FPP DICOM, which we then used to fine-tune our model for FPP reconstruction (Fig. \ref{fig:transfer_method}).



\subsection{Dataset and Implementation Details}

Clinical cine and FPP 1.5T CMR datasets were used in this study. Data were collected using clinical protocols from volunteers and patients with local IRB approval. 
The original cine data consisted of 418 single-slices from 123 patients, resulting in a total of 3294 SMS data using intra-patient slice combinations. The dataset was split 3130/164 for training/testing. 
For FPP training/validation, we used 82 DICOM slices from 27 patients to simulate 1485 SMS data. For testing, 
19 slices from 6 patients generated 21 undersampled SMS data.
For all experiments, in-plane acceleration rate of 2 and SMS acceleration rate of 2 were simulated, amounting to a total acceleration rate of 4. 

The CRNN model performed 2D convolution and feature aggregation in 3 different directions: layer, time frame, and iteration. Similar structure has been used in other single-slice dMRI reconstruction \cite{chen2020real,qin2018convolutional}. A 3D convolution in 2D + time directions was added right before DC. We performed 5 iterations of the network forward pass and DC steps (Fig. \ref{fig:recon_outline}). For the training loss, we used a weighted sum of mean squared error (MSE), structural similarity index (SSIM), and the temporal total variation (TV) loss, with the weights determined via grid search. The results were as expected for hyperparameter changes within one order of magnitude. We used the Adam optimizer \cite{kingma2014adam} with a fixed learning rate of 0.0001. Experiments were performed using PyTorch v1.10 on NVIDIA A100 GPU. Inference required $\sim$ 4Gb of memory.

\section{Experiments and Results}

\begin{table} [t] \label{table:results}
\def\sym#1{\ifmmode^{#1}\else\(^{#1}\)\fi}
\sisetup{detect-weight,mode=text}
\robustify\bfseries
\caption{Network architecture comparisons (top) and transfer learning results (bottom). * signifies statistical significance (paired student's t-test, $p<0.05$) compared to all other experimental conditions. Fail $\coloneqq$ NMSE$>$1000.}
\centering
\begin{tabular*}{\textwidth}{ l @{\extracolsep{\fill}} S[table-format=2.1(4),separate-uncertainty=true,table-align-text-post = false, table-space-text-post=\sym{*\dagger}]
S[table-format=2.1(4),separate-uncertainty=true,table-align-text-post = false, table-space-text-post=\sym{*\dagger}]
S[table-format=1.3(4),separate-uncertainty=true,table-align-text-post = false, table-space-text-post=\sym{*\dagger}]
S[table-format=2,separate-uncertainty=true,table-align-text-post = false]
S[table-format=2,separate-uncertainty=true,table-align-text-post = false]
}
\toprule

& {\makecell{NMSE  \\ $\times 10^{-3} \downarrow$}} & {\makecell{PSNR $\uparrow$}} & {\makecell{SSIM $\uparrow$}} &{\makecell{Fail \\ (\%)}} &{\makecell{Recon \\ time (s)}} \\

\midrule

Independent slices
& 55.0 \pm 21.9 & 31.0 \pm 2.4 & 0.834 \pm 0.040 & 0  & 1.63 \\

3D convolution
& 4.5 \pm 2.2 & 42.0 \pm 2.6 & 0.972 \pm 0.008 & 0 & 8.49 \\

\bfseries Proposed network
& \bfseries 1.9 \pm 1.0 \sym{*} & \bfseries 45.8 \pm 2.7 \sym{*} & \bfseries 0.988 \pm 0.005 \sym{*} & 0 & 1.77 \\

\midrule

No FPP fine-tune
& 48.2 \pm 86.2 & 35.1 \pm 4.1 & 0.939 \pm 0.017 & 14 & 0.93 \\

No cine pre-train
& 10.1 \pm 4.3 & 39.2 \pm 2.0 & 0.954 \pm 0.014 & 0 & 0.95 \\

Simple phase
& 9.8 \pm 3.9 & 39.3 \pm 1.8 & 0.951 \pm 0.013 & 0 & 0.96 \\

\bfseries Proposed transfer
& \bfseries 7.8 \pm 3.1 \sym{*} & \bfseries 40.3 \pm 1.7 \sym{*} & \bfseries 0.960 \pm 0.012 \sym{*} & 0 & 0.95 \\

\bottomrule
\end{tabular*}
\end{table}

\begin{figure}[t] 
  \centering
  \centerline{\includegraphics[width=11.5cm]{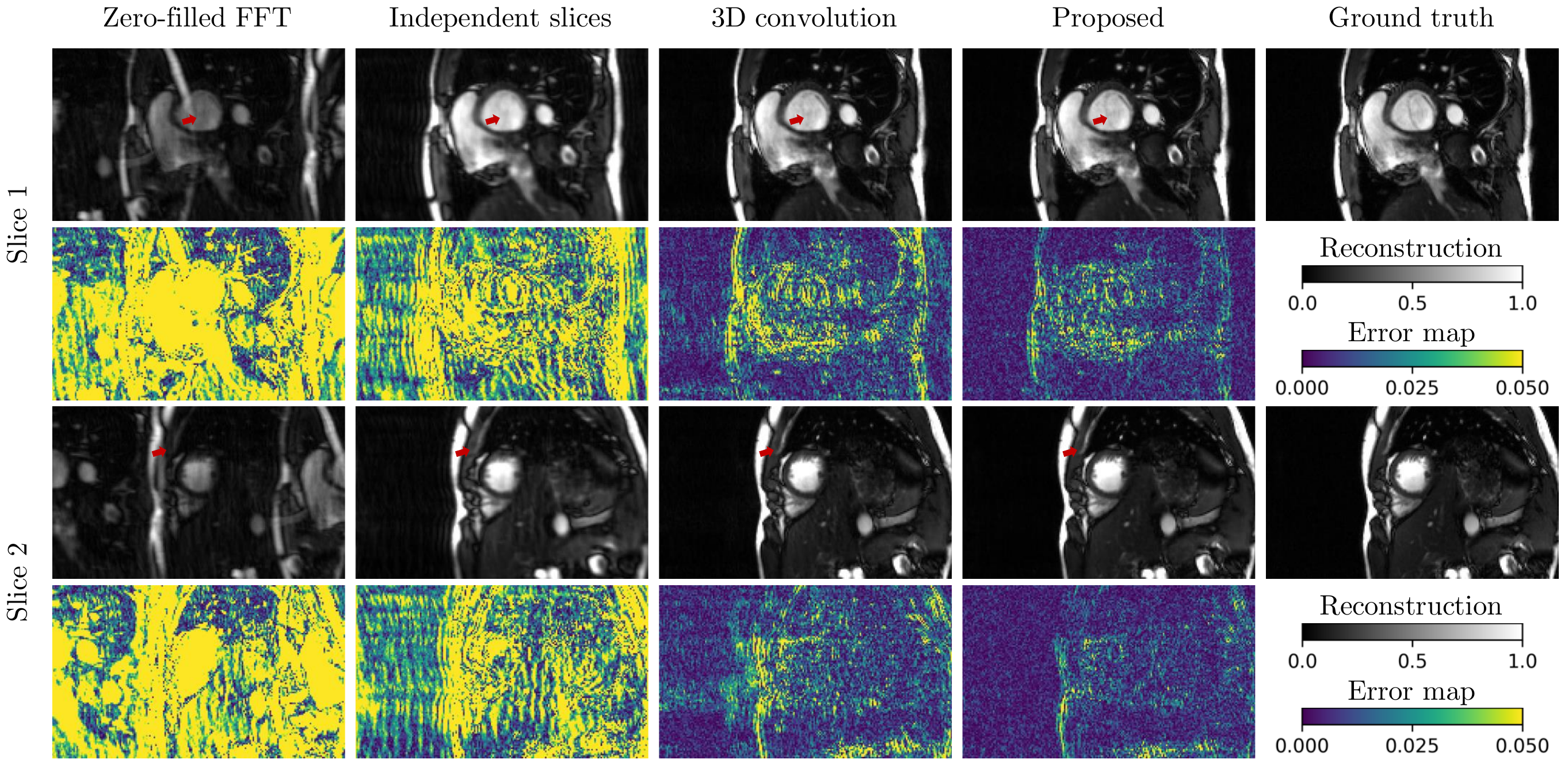}} 
  \caption{Cine image slices for different dynamic SMS reconstruction methods.}
  \label{fig:network_comparisons}
\end{figure}

\subsection{Dynamic SMS Reconstruction Strategy Comparison}

We compared our results to two experimental conditions specifically focusing on the method of handling SMS slices: (1) ``Independent slices", where each undersampled SMS image slice was treated as a separate training example for a single CRNN, and (2) ``3D convolution", where convolutions in the CRNN were replaced with 3D convolutions in the spatial and slice ($(a,b,i)$) dimensions. The independent-slice method can be thought of as a naive application of previous CRNN methods \cite{chen2020real,qin2018convolutional}, but without the DC layer because the original SMS k-space input is not compatible with the k-space of a single-slice image. The 3D convolution method is another possible way for the network to use information from multiple SMS slices.

As shown in Table \ref{table:results}, our weight-sharing CRNN outperforms both conditions by a large margin in all common image quality metrics - normalized mean square error (NMSE), peak signal-to-noise-ratio (PSNR), and structural similarity index (SSIM). The high quality of our final reconstruction shown in Fig. \ref{fig:network_comparisons} corroborate these metrics, and the error maps help visualize the improvements.

Compared to 3D convolution, the second-best performing method, our proposed method still removes a considerable amount of artifacts. One potential explanation for this difference is that a 3D convolution in spatial and slice directions would result in inappropriate aggregation of unrelated inter-slice information, since the slices are far apart and inter-slice artifacts are translated significantly from the original image. Our weight-sharing scheme instead allows for sharing of local denoising filters between the SMS slices, which is less dependent on features aligning in the inter-slice direction.

\subsection{Transfer Learning for FPP Reconstruction}

\begin{figure}[t] 
  \centering
  \centerline{\includegraphics[width=11.5cm]{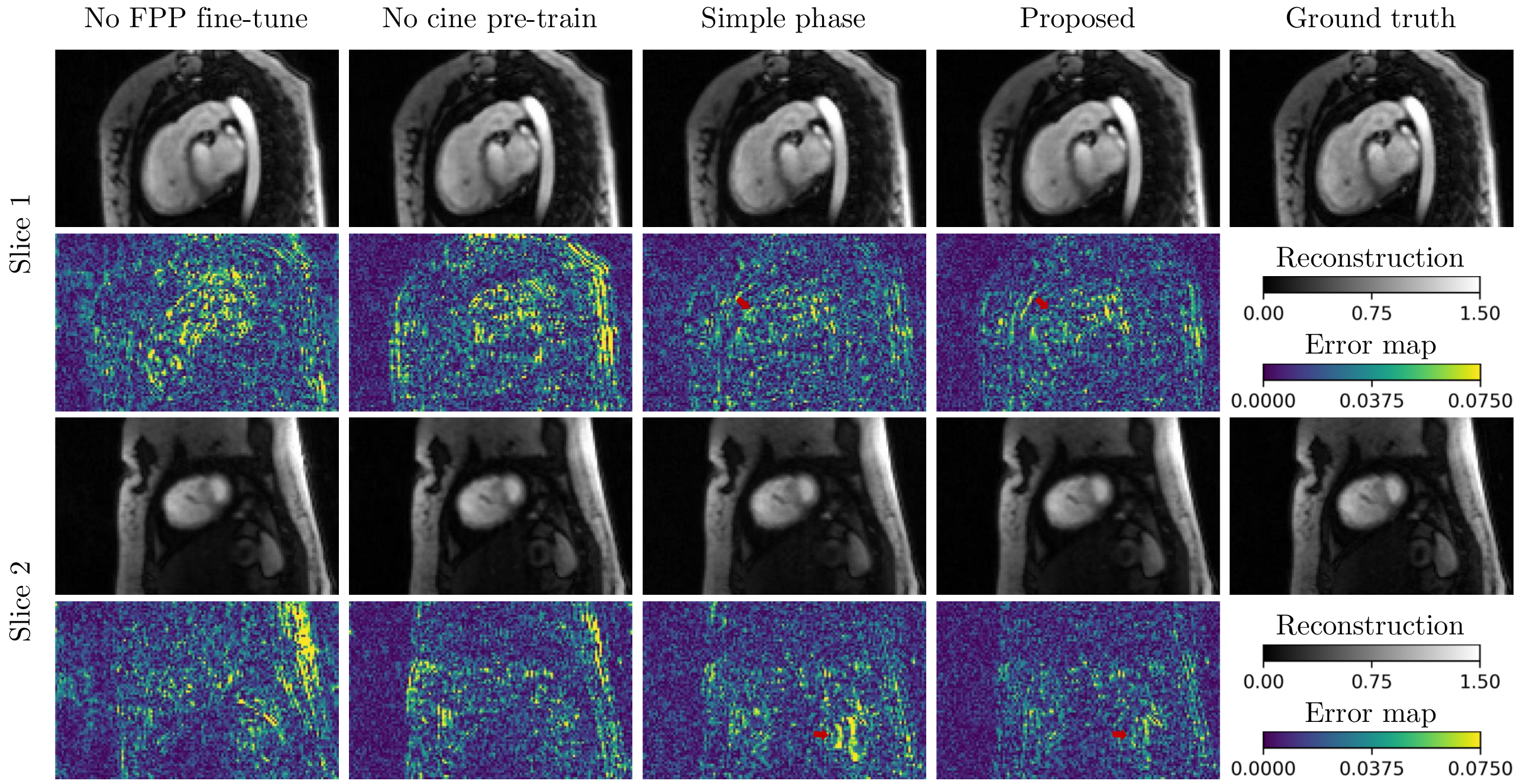}}
  \caption{FPP image slices for different transfer learning methods.}
  \label{fig:transfer_results}
\end{figure}

We compared our proposed transfer learning strategy to three other conditions: (1) ``no FPP fine-tune" - trained only with cine, (2) ``no cine pre-train" - trained from scratch using the proposed synthetic FPP data, and (3) ``simple phase" - pre-trained with cine and fine-tuned with simplified synthetic FPP data with sinusoidal grating phases (i.e. no phase changes based on the image content). As shown in Table \ref{table:results}, our proposed method outperforms all other methods for all metrics with statistical significance. This difference in performance can also be visualized by the error maps in Fig. \ref{fig:transfer_results}, which show both increased reduction in fuzzy noises as well as better accuracy in intensity estimation in regions of interest (e.g. the left and right ventricles). These results validate the importance of all of our proposed transfer learning steps in Fig. \ref{fig:transfer_method}.

\section{Discussions and Conclusion} \label{sec:future_works}

The main limitation of our work is in the handcrafted nature of the synthetic phases and CSMs. Since we have confirmed that good synthetic phases and CSMs are important for successful transfer learning, an interesting future direction could be to use the cine k-space dataset to learn a generative model that can convert FPP DICOM data to synthetic multi-coil k-space data. Furthermore, we could extend our analysis beyond CMR and validate our methods' generality by performing evaluations on other dynamic SMS MRI datasets.

In this work, we presented (1) a novel framework for incorporating inter-slice dependencies in deep learning-based dynamic SMS MRI reconstruction and (2) an MR-based transfer learning strategy for limited-data FPP reconstruction. Our proposed methods show improved performance over existing approaches and produce high quality reconstructions for both cine and FPP imaging.

\setcounter{secnumdepth}{2}






\bibliographystyle{splncs04}
\bibliography{refs}
\end{document}


\begin{itemize}
\item[$\blacksquare$] Proof for 3D conversion of SMS reconstruction
\end{itemize}
Let $I(x,y,z)$ and $S(k_x,k_y,z)$ be the fully sampled image (object) and the corresponding k-space signal collected at slice location $z$, respectively. $S(k_x,k_y,z) = FFT_{xy}(I(x,y,z))$, where $FFT_{xy}$ represents the 2D Fourier transform along $x$ and $y$ directions. The SMS k-space signal is a summation of the signals from all simultaneously excited slices $S_{sms}(k_x,k_y) = \sum_{z}{S(k_x,k_y,z)}$.

By manipulating the excitation radio frequency pulse, an additional phase can be added to the signal. Without loss of generality, assume the k-space is collected on a Cartesian trajectory and line-by-line. If each k-space readout line from slice $z$ has an accumulating phase of $2 \pi z/M$, where $M$ is the number of SMS slices, then 
\begin{equation}
S_{sms}(k_x, k_y)=\sum_{z}{S(k_x,k_y,z)e^{j2 \pi/M z k_y}}
\end{equation}
Using the property that $e^{j2 \pi i} = e^{j2 \pi i + 2 \pi}$, we can define $k_z := k_y \bmod{M}$, and
\begin{equation}
S_{sms}(k_x, k_y)=\sum_{z}{S(k_x,k_y,z)e^{j2 \pi/M z k_z}}
\end{equation}
Exploit the definition of Fourier transform and treat $k_z$ as a new dimension, $S_{sms}(k_x, k_y)$ is converted to a 3D tensor and 
\begin{equation}
\begin{split}
S_{sms}(k_x, k_y, k_z) & = \sum_{z}{S(k_x,k_y,z)e^{j2 \pi/M z k_z}} \\
& = FFT_{z}(S(k_x,k_y,z)) \\
& = FFT_{xyz}(I(x,y,z))
\end{split}
\end{equation}

\newpage

\begin{itemize}
\item[$\blacksquare$] Details of the CRNN module
\end{itemize}
\begin{figure}[H]
  \centering
  \centerline{\includegraphics[width=11.5cm]{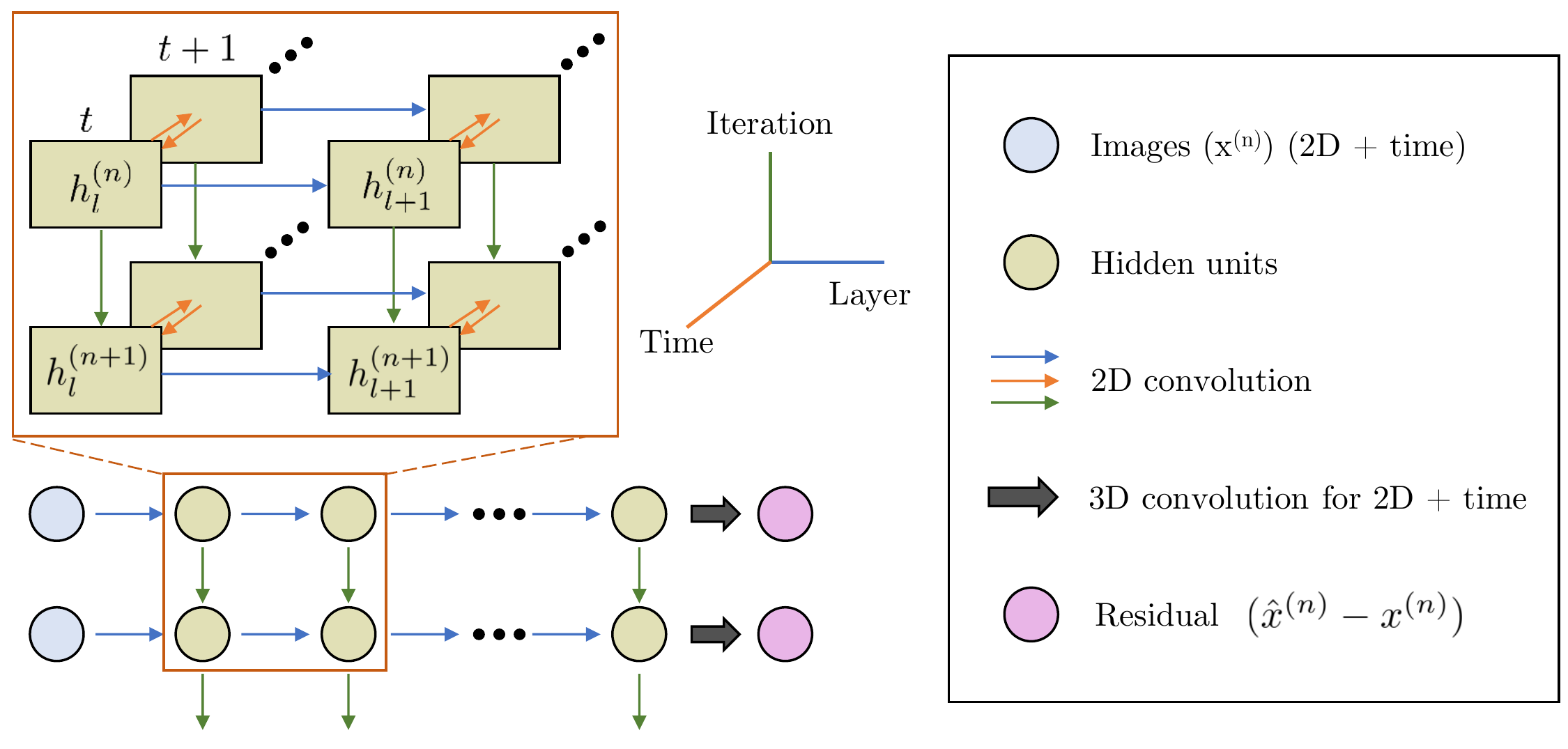}}
  \caption{Details of the CRNN layers, where $h_l^{(n)}$ is the hidden units at layer $l$ at iteration $n$, and $t$ is the time index for the image sequence. For brevity, the SMS slice index is omitted. Multiple arrows are combined using element-wise summation followed by a non-linearity.
  }
\end{figure}

\begin{itemize}
\item[$\blacksquare$] Evidence for domain gap caused by value type and coil configuration
\end{itemize}
\begin{table}
\def\sym#1{\ifmmode^{#1}\else\(^{#1}\)\fi}
\sisetup{detect-weight,mode=text}
\robustify\bfseries
\caption{Evidence that phase and CSM significantly contribute to the domain gap between the training and testing data. Comparing second and third rows, noticeable improvement was made by simply applying similar synthetic phase and CSM steps to the training and testing data. This experiment is merely to confirm the importance of phase and CSM; we cannot alter the phase and CSM of raw k-space data to match our simulation. Therefore, we instead try to match our estimation of phase and CSM as closely as possible to the real phase and CSM.}
\centering
\begin{tabular*}{\textwidth}{ l @{\extracolsep{\fill}} l  S[table-format=1.3,separate-uncertainty=true,table-align-text-post = false, table-space-text-post=\sym{*\dagger}]
S[table-format=2.1,separate-uncertainty=true,table-align-text-post = false, table-space-text-post=\sym{*\dagger}]
S[table-format=1.3,separate-uncertainty=true,table-align-text-post = false, table-space-text-post=\sym{*\dagger}]
}
\toprule

Training condition & Test data & {\makecell{NMSE $\downarrow$}} & {\makecell{PSNR $\uparrow$}} & {\makecell{SSIM $\uparrow$}} \\

\midrule
\midrule

Train with Cine & Raw FPP k-space & 0.266 & 32.9 & 0.930 \\

\midrule

\makecell[l]{Pre-train with Cine \\ $\rightarrow$ Fine-tune with \\ Synthetic FPP \\ with simple phase} & Raw FPP k-space
& 0.010 & 38.7 & 0.948 \\

\midrule

\makecell[l]{Pre-train with Cine \\ $\rightarrow$ Fine-tune with \\ Synthetic FPP \\ with simple phase} & \makecell[l]{Raw FPP k-space \\ $\rightarrow$ convert to DICOM \\ $\rightarrow$ \makecell[l]{Synthetic FPP \\ with simple phase}}
& \bfseries 0.003 & \bfseries 43.5 & \bfseries 0.982 \\

\bottomrule
\end{tabular*}
\end{table}